\pdfoutput=1
\RequirePackage{fix-cm}
\documentclass[smallextended]{svjour3}       
\smartqed  
\usepackage{graphicx}

\usepackage{amsmath}
\usepackage{rotate}
\usepackage{colortbl}
\usepackage{arydshln}
\usepackage{times}

\usepackage[dvipsnames]{xcolor}
\usepackage{rotating}
\usepackage{makecell} 
\usepackage{tabularx}
\usepackage{balance}
\usepackage{booktabs}
\usepackage{wrapfig}
\usepackage{tikz}
\usetikzlibrary{angles}
\usepackage{makecell}
\usepackage{tabu}
\usepackage{multirow}

\usepackage[numbers]{natbib}

\newcommand{\bi}{\begin{itemize}}
\newcommand{\ei}{\end{itemize}}
\newcommand{\be}{\begin{enumerate}}
\newcommand{\ee}{\end{enumerate}}

\newcommand{\tbl}[1]{Table~\ref{tbl:#1}}
\newcommand{\fig}[1]{Figure~\ref{fig:#1}}

\usepackage{url}

\newcommand{\crule}[3][darkgray]{\textcolor{#1}{\rule{#2}{#3}}}

\newcommand{\dbox}[1] { \crule[black!#1]{0.67cm}{0.4cm} 
\hspace{-0.51cm}\scalebox{1}[1.0]{{\textcolor{black}{{\bf $^{#1}$}}}\hspace{0.1mm}}}

\newcommand{\wbox}[1] { \crule[black!#1]{0.67cm}{0.4cm} 
\hspace{-0.51cm}\scalebox{1}[1.0]{{\textcolor{white}{{\bf $^{#1}$}}}\hspace{0.1mm}}}

\newcommand{\quart}[4]{
\begin{picture}(100,6)
    {
        \color{black}
        \put(#3,3)
        {\circle*{4}}
        \put(#1,3)
        {\line(1,0){#2}}
    }
\end{picture}
}

\newcommand{\ofr} {
{\textit{out-of-range}}
}

\usepackage[framemethod=tikz]{mdframed}
\usetikzlibrary{shadows}
\usepackage{graphics}
\newmdenv[
tikzsetting= {fill=blue!10},
linewidth=1pt,
roundcorner=2pt, 
shadow=false
]{myshadowbox}

\newenvironment{result}[2]
{\begin{myshadowbox}\textbf{\textit{\underline{Lesson#1:}}} #2}{ 
\end{myshadowbox}}
    
\newcommand{\nm}[1]{\hline\multicolumn{1}{c}{\cellcolor{black} { {\bf \textcolor{white}{#1}}}}}
\begin{document}

\title{Hyperparameter Optimization  for    Effort Estimation
}


\newcommand{\RED}{\color{black}}
\newcommand{\BLACK}{\color{black}}

\author{Tianpei Xia         \and
        Rahul Krishna         \and
        Jianfeng Chen         \and
        George Mathew         \and
        Xipeng Shen         \and
        Tim Menzies         
}


\institute{Tianpei Xia, Rahul Krishna, Jianfeng Chen, George Mathew, Xipeng Shen,
Tim Menzies\at
              Department of Computer Science, North Carolina State University, Raleigh, NC, USA \\
              Email: txia4@ncsu.edu,  rkrish11@ncsu.edu, jchen37@ncsu.edu, george2@ncsu.edu, xshen5@ncsu.edu, timm@ieee.org 
}

\date{the date of receipt and acceptance should be inserted later}

\maketitle

\begin{abstract}
\RED
Software analytics has  been widely used in software engineering for many tasks. One of the ``black arts'' of software analytics is tuning the 
parameters controlling a machine learning algorithm.
Such a {\em hyperparameter optimization} has been found to
be very  useful   in SE defect prediction and text mining~\cite{agrawal2017better,fu2016differential,Fu2016TuningFS,menzies2018500+,atkinson2003semantically}.
Accordingly, this paper seeks simple (easy to implement), automatic (no manually hyperparameter settings), effective (high performance in experiment) and fast (Running time is fast) methods
\BLACK for finding   tunings for   software effort estimation.

We introduce a     hyperparameter optimization architecture called
  OIL  (Optimized Inductive Learning). 
  OIL is tested on a wide range of    optimizers using  
  data from 945  projects. After tuning,
  large improvements in effort estimation accuracy were observed
  (measured in terms of magnitude of relative error and standardized accuracy).
  
  From those results, we recommend
  using  regression trees (CART) tuned by either different evolution or FLASH (a
  sequential model optimizer).  This   particular combination of learner and optimizers   achieves superior results
in a few minutes (rather than the many hours required by some other approaches).

\end{abstract}

\section{Introduction}
\label{intro}
Software analytics has  been widely used in software engineering for many tasks~\cite{menzies2018software}. 
This paper explores methods to improve algorithms
for software effort estimation (a particular kind of analytics tasks).
This is needed since software  effort estimates   can be 
wildly inaccurate~\cite{kemerer1987empirical}. Effort estimations need to be accurate  (if for
no other reason) since many government
organizations demand that the budgets allocated to
large publicly funded projects be double-checked by some estimation model~\cite{MenziesNeg:2017}.  
Non-algorithm techniques that rely on human judgment~\cite{jorgensen2004review} are much harder to  
audit or 
dispute (e.g.,  when the estimate is generated by a senior colleague but disputed by others). 

Sarro et al.~\cite{sarro2016multi} assert that effort  estimation  is  a  critical  activity  for  planning  and
monitoring software project development in order to deliver
the  product  on  time  and  within  budget~\cite{briand2002resource,kocaguneli2011experiences,trendowicz2014software}.   The
competitiveness of software
organizations depends on their ability to accurately predict
the  effort  required  for  developing  software  systems;  both
over- or under- estimates can negatively affect the outcome
of software projects~\cite{trendowicz2014software,mcconnell2006software,mendes2002further,sommerville2010software}.

{\em Hyperparameter optimizers}   tuning
 the control parameters of a data
mining algorithm.
It is well established that  classification tasks like software defect prediction or 
 text classification are improved by such tuning~\cite{Fu2016TuningFS,tanti18,AGRAWAL2018,agrawal2017better}.
This paper investigates hyperparameter optimization using data from 945   projects, the study is an extensive exploration  of
hyperparameter optimization and effort estimation \RED following the earlier work done by Corazza et al~\cite{corazza2013using}.
\BLACK

We assess our results with respect to recent findings by 
Arcuri \& Fraser~\cite{Arcuri2013}.
They caution that to transition hyperparameter optimizers  to industry,   they need to be {\bf  fast}:
\begin{quote}
{\em  
A practitioner, that wants to use such tools, should not be required to run large tuning phases before being able to apply those tools~\cite{Arcuri2013}.}
\end{quote}
Also,  according to Arcuri \& Fraser, optimizers must  be {\bf useful}:
\begin{quote}
{\em At least in the context of test data generation, (tuning) does not seem easy to find good settings that significantly outperform   ``default'' values. ... Using ``default'' values is a reasonable and justified choice, whereas parameter tuning is a long and expensive process that might or might not pay off~\cite{Arcuri2013}.}
\end{quote}
Hence, to  assess  such optimization for effort estimation, we ask four  questions.

{\bf RQ1:} To address one concern raised by Arcuri \& Fraser,
 we must  first ask {\em is it best   to just  use ``off-the-shelf'' defaults?}
We will find that  tuned learners
provide  better estimates  than untuned learners. Hence, for effort estimation:
 \begin{result}{1}
 ``off-the-shelf'' defaults should be deprecated.
 \end{result}

{\bf RQ2:} 
  Can tuning effort be avoided by   {\em  replacing  old defaults with new defaults}?
 This  checks if we can run tuning once (and once only) then use those new defaults
 ever after. We will observe that effort estimation tunings differ extensively
 from dataset to dataset. Hence, for effort estimation:
 \begin{result}{2}
 Overall, there are no ``best'' default settings.
 \end{result}

{\bf RQ3:}  The first two research questions tell us that we must retune
our effort estimators whenever new data arrives. Accordingly, we must now address the other concern raised by Arcuri \& Fraser
about CPU cost. Hence, in this question we ask   {\em can we avoid  
slow hyperparameter optimization}? The answer to  {\bf RQ3} will be ``yes'' since
our results show that for effort estimation:
 \begin{result}{3}
 Overall, our slowest optimizers  perform no better than  faster ones.
 \end{result}

{\bf RQ4:} The
final question to  answer is {\em what  hyperparameter   optimizers  to  use  for  effort
estimation?} Here, we report that a certain combination  of learners and optimizers usually
produce best results. Further, this
particular combination  often achieves in a few
minutes what other optimizers may need hours to days of CPU to achieve. Hence we will recommend 
the following combination for effort estimation:
\begin{result}{4}
For new datasets, try a combination of {\em CART} with the optimizers {\em differential evolution} and {\em FLASH}.
\end{result}
(Note: The {\it italicized} words are explained below.)

In summary, unlike the test case generation domains explored by Arcuri \& Fraser,
hyperparamter optimization for effort estimation is both {\bf useful} and {\bf fast}.

Overall the contributions of this paper are:
\bi
\item A demonstration that defaults settings are not the best way to perform effort estimation. Hence, when new
data is encountered, some tuning process is required
to learn the best settings for generating estimates
from that data.
\item A recognition of the inherent difficulty associated
with effort estimation.  Since there is not one universally best effort estimation method.
commissioning a new effort estimator requires
extensive testing.  
As shown below, this can take  hours to days of CPU time. 
\item
The identification of a  combination
of learner and optimizer that  works
as well as  anything else, and which takes minutes to learn an effort estimator.
\item An extensible open-source architecture called OIL that enables the commissioning of effort estimation methods.
OIL makes our results   repeatable and refutable.
\ei
The rest of this paper is structured as follows.
The next section discusses different methods for effort estimation and how to optimize
the parameters of effort estimation methods. This is followed by a description of our data, our experimental
methods, and our results. After that, a discussion section explores open issues with this work. 

From all of the above, we can conclude  that (a)~   Arcuri \& Fraser's pessimism about hyperparameter optimization  applies to their
test data generation domain. \RED However (b)~for   effort estimation,  hyperparamter optimization is both {\bf useful} and {\bf fast}. \BLACK
Hence, we hope that OIL, and the results of  this paper,  will prompt and enable   more research on   methods
to tune software effort estimators.

Note that OIL and all the data used in this study is freely available for download from https://github.com/arennax/effort\_oil\_2019

\section{About Effort Estimation}
\label{sec:1}
Software effort estimation is \RED a method to offer managers
approximate advice on how much  human effort (usually expressed in terms of hours, days or months of human work) is required to plan, design and develop a software project.  \ Such advice
can only ever be approximate due to dynamic nature of any software development.  Nevertheless, 
it is important to attempt \BLACK to allocate resources properly in software projects to avoid waste. In some cases, inadequate or overfull funding can cause a considerable waste of resource and time~\cite{cowing02,germano16,hazrati11,roman16}. 
As shown below, effort estimation can be categorized into (a)~human-based and (b)~algorithm-based methods~\cite{teak2012,shepperd2007software}. 

For several reasons, this paper does not explore human-based  estimation methods.
Firstly,   it is known that humans rarely update their human-based estimation knowledge
based on feedback from new projects~\cite{jorgensen2009impact}.
Secondly, algorithm-based methods are preferred when   estimate have to be audited or debated
(since the  method is explicit and   available for inspection). 
Thirdly, algorithm-based methods can be run many times (each time applying small mutations to the input data)
to understand the range of possible estimates.
Even very strong
advocates of human-based methods~\cite{jorg2015a} acknowledge that algorithm-based methods are useful for learning
the
uncertainty
about particular estimates.

\subsection{Algorithm-based Methods}
\label{sec:3}

There are many  algorithmic estimation methods. Some, such as COCOMO~\cite{boehm1981software}, make assumptions
about the  attributes in the model. For example, COCOMO requires that 
data includes 22 specific   attributes such as analyst capability (acap) and software complexity (cplx).  This   attribute
assumptions restricts how much data is available for studies like this paper. 
For example,
here we explore 945 projects expressed using a  wide range of  attributes.  If we used COCOMO, we could only have accessed an order of magnitude   fewer projects. 

Due to its attribute assumptions, this paper does not study COCOMO data. 
All the following learners can accept projects described using any attributes, just as long as one of those is some measure of project development effort.


Whigham et al.'s ATLM method~\cite{Whigham:2015} is a multiple linear regression model which calculate the effort as $\mathit{effort} = \beta_0 + \sum_i\beta_i\times a_{i} +  \varepsilon_i$,  where $a_i$ are explanatory attributes and $\varepsilon_i$ are errors to the actual value. The prediction weights $\beta_i$ are determined using least square error estimation~\cite{neter1996applied}. Additionally, transformations are applied on the attributes to further minimize the error in the model. In case of categorical attributes the standard approach of ``dummy variables"~\cite{hardy1993regression} is applied. While, for continuous attributes, transformations such as logarithmic, square root,  or no transformation is employed such that the skewness of the attribute is minimum. It should be noted that, ATLM does not consider relatively complex techniques like using model residuals,  box transformations or step-wise regression (which are standard) when developing a linear regression model. The authors make this decision since they intend ATLM to be a simple baseline model rather than the ``best" model. \RED And since it can be applied automatically, there should be no excuse not to compare any new model against a comparatively naive baseline.

\BLACK
Sarro et al. proposed a method named Linear Programming for Effort Estimation (LP4EE)~\cite{SarroTOSEM2018}, which aims to achieve the best outcome from a mathematical
model with a linear objective function subject to linear equality and inequality
constraints. The feasible region is given by the intersection of the constraints and the
Simplex (linear programming algorithm) is able to find a point in the polyhedron where
the function has the smallest error in polynomial time. In effort estimation problem, this model minimizes the Sum of Absolute Residual (SAR), when a new project is presented to the model, LP4EE predicts the effort as $\mathit{effort} = a_1*x_1 + a_2*x_2 + ... + a_n*x_n$, where $x$ is the value of given project feature and $a$ is the corresponding coefficient evaluated by linear programming.
LP4EE is suggested to be used as another baseline model for effort estimation since it provides
similar or more accurate estimates than ATLM and is much less sensitive than ATLM
to multiple data splits and different cross-validation methods.

Some algorithm-based estimators are regression trees such as   CART~\cite{brieman84}, 
\BLACK CART is a  tree learner that divides a dataset, then recurses
on each split.
If data contains more than {\em min\_sample\_split}, then a split is attempted.
On the other hand, if a split contains no more than {\em min\_samples\_leaf}, then the recursion stops. 
CART finds the attributes whose ranges contain rows with least variance in the number
of defects. If an  attribute ranges $r_i$ is found in 
$n_i$ rows each with an effort  variance of $v_i$, then CART seeks the  attribute with a split that most
minimizes $\sum_i \left(\sqrt{v_i}\times n_i/(\sum_i n_i)\right)$.
For more details on the CART parameters, see Table~\ref{tbl:cart}.

\begin{table*}[!t]
\centering
\caption{CART's parameters.}\label{tbl:cart}
\begin{tabular}{r|c|c|p{2.26in}}
\hline
Parameter          & Default & Tuning Range  & Notes                                                            \\\hline
max\_feature       & None    & {[}0.01, 1{]} & The number of feature to consider when looking for the best split. \\\hline 
max\_depth         & None    & {[}1, 12{]}   & The maximum depth of the tree.                                     \\\hline 
min\_sample\_split & 2       & {[}0, 20{]}   & Minimum   samples required to split  internal nodes.  \\\hline 
min\_samples\_leaf & 1       & {[}1, 12{]}   &   Minimum   samples required to be at a leaf node.    \\\hline  

\end{tabular}
\end{table*}

\RED Random Forest~\cite{breiman2001random} and Support Vector Regression~\cite{chang2011libsvm} are another instances of regression methods. Random Forest (RF) is an ensemble learning method for  regression (and classification) tasks that  builds a set of   trees when training the model. To decide the output, it uses 
the mode of the classes (classification) or mean prediction (regression) of the individual trees.
Support Vector Regression (SVR) uses kernel functions to project
the data onto a new hyperspace where complex non-linear patterns
can be simply represented. It aims to construct an optimal hyperplane that fits data and predicts with
minimal empirical risk and complexity of the modelling function.

\begin{figure}
\vspace{0.5cm}
\centerline{\includegraphics[width=1\textwidth]{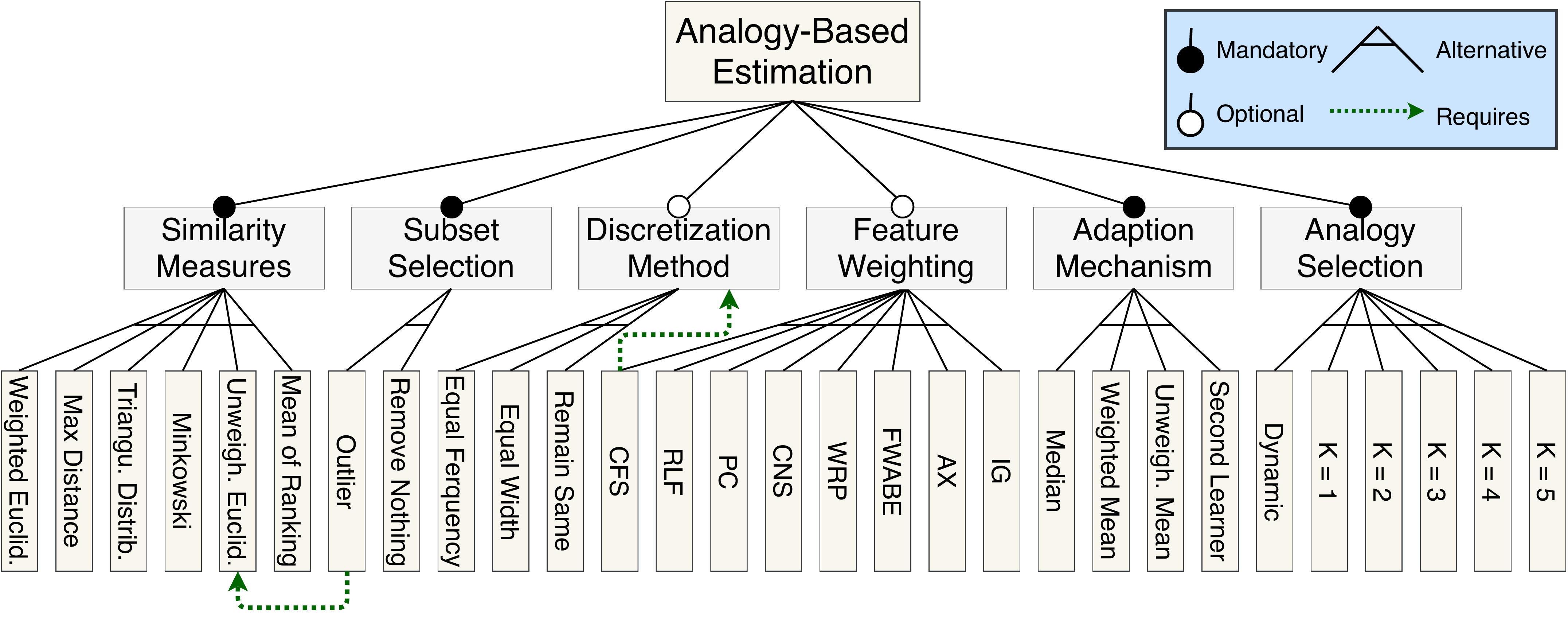}}
\caption{OIL's feature model of the space of machine learning options for ABEN.  In this model, $\mathit{Subset Selection}$, $\mathit{Similarity}$, $\mathit{Adaption Mechanism}$ and $\mathit{Analogy Selection}$ are the mandatory {\em features}, while the $\mathit{Feature Weighting}$ and $\mathit{Discretization Method}$ {\em features} are optional. To avoid making the graph too complex, some cross-tree constrains are not presented. For more details on the terminology of this figure, see  \tbl{aben}.}    
\label{fig:featuretree}
\end{figure}

\BLACK Another algorithm-based estimators are the 
analogy-based Estimation (ABE) methods advocated by Shepperd and Schofield~\cite{shepperd1997estimating}. ABE is widely-used~\cite{7194627,Kocaguneli2015,7426628,6092574,MenziesNeg:2017}, in many forms.
We  say that  ``ABE0'' is the standard  form  seen in the literature
and ``ABEN'' are the 6,000+ variants of ABE  defined below. 
The general form of ABE (which applies to  ABE0 or ABEN) is
to first form a table of rows of past projects. The {\em columns} of this table are composed of independent variables (the {\em features} that define projects) and one dependent {\em feature} (project  effort).
From this table, we learn  what  similar projects (analogies) to use from the training set when examining a new test instance.
For each test instance, ABE then selects   $k$ analogies out of the training set.
Analogies are selected via   a {\em similarity measure}. 
Before calculating similarity,  ABE normalizes   numerics  min..max to 0..1 (so all numerics   get equal chance to influence the dependent). 
Then, ABE uses {\em feature} weighting to reduce the influence of less informative {\em features}.
Finally, some  {\em adaption strategy} is applied  return a combination of  the dependent effort values seen in  the $k$ nearest analogies. For  details on ABE0 and ABEN, see \fig{featuretree} \& Table~\ref{tbl:aben}.

\begin{table}
\caption{Variations on analogy. 
 Visualized in
\fig{featuretree}.}\label{tbl:aben}

\begin{tabular}{|p{.94\linewidth}|}\hline
\small
\begin{itemize}
\item
To measure    similarity between  $x,y$, 
ABE uses $\sqrt{\sum_{i=1}^n w_i(x_i-y_i)^2}$ where  $w_i$ corresponds to {\em feature} weights applied to independent {\em features}. ABE0 uses a uniform weighting where $w_i=1$.
ABE0's {\em adaptation strategy} is to return the  effort   of the nearest $k=1$   item.
\item
{\em Two ways to find training subsets}:
(a) Remove nothing: Usually, effort estimators use all training projects~\cite{chang1974finding}. Our ABE0 is using this variant;
(b) Outlier methods: prune training projects with (say) suspiciously large  values~\cite{keung2008analogy}. 
\item
{\em Eight ways to make feature weighting}:
Li {\it et al.}~\cite{li2009study} and Hall and Holmes~\cite{hall2003benchmarking} review 8 different {\em feature} weighting schemes.
\item
{\em Three ways to discretize} (summarize numeric ranges into a few bins):
Some {\em feature} weighting schemes require an initial discretization of continuous  columns. There are many discretization policies in the literature, including:
(1)~equal frequency,
(2)~equal width, 
(3)~do nothing.
\item
{\em Six ways to choose similarity measurements:}
Mendes {\it et al.}~\cite{mendes2003comparative} discuss three similarity measures, including the weighted Euclidean measure described above, an unweighted variant (where $w_i$ = 1), and a ``maximum distance'' measure that focuses on the single {\em feature} that maximizes interproject distance. Frank {\it et al.}~\cite{frank2002locally} use a triangular distribution that sets to the weight to zero after the distance is more than ``k'' neighbors away from the test instance. A fifth and sixth similarity measure are the Minkowski distance measure used in~\cite{angelis2000simulation} and the mean value of the ranking of each project {\em feature} used in~\cite{walkerden1999empirical}.
\item
{\em Four ways for adaption mechanisms:} 
(1)~median effort value,
(2)~mean dependent value,
(3)~summarize the adaptations via a second learner (e.g., linear regression)~\cite{li2009study,menzies2006selecting,baker2007hybrid,quinlan1992learning},
(4)~weighted mean~\cite{mendes2003comparative}.
\item
{\em Six ways to select analogies:}
Analogy selectors  are  fixed or dynamic~\cite{teak2012}. Fixed methods use $k\in\{1,2,3,4,5\}$
nearest neighbors
while  dynamic methods use the training set to find which $1 \le k \le N-1$ is best for   $N$ examples.
\end{itemize}
\\\hline
\end{tabular}
\end{table}

\subsection{Effort Estimation and  Hyperparameter Optimization}
\label{sec:4}

\RED Note that we do \underline{{\em not}} claim that the above represents all methods for  effort estimation. Rather, we  say that (a)~all the above are either prominent in the literature or widely used; and (b)~anyone  with knowledge of  the current effort estimation 
literature would be tempted to try some of the above.
\BLACK

Even though our lost of effort estimation methods is incomplete, it is still
 very long. Consider, for example, just the ABEN
variants documented in Table~\ref{tbl:aben}. There are 
$2\times 8\times 3\times 6\times 4\times 6=6,912$ such variants.  Some   can be ignored;
e.g. at $k=1$,     adaptation mechanisms return the same result, so they are not necessary. Also, not all  {\em feature} weighting techniques use discretization. But even after those discards, there are still thousands of possibilities. 

Given the space to exploration is so large, some researchers have offered automatic support for that exploration.
Some of that prior work suffered from being applied to limited data~\cite{li2009study}.



Other researchers assume that the effort model is a specific parametric form (e.g. the COCOMO equation)
and propose mutation methods to adjust the parameters of that equation~\cite{aljahdali2010software,Moeyersoms:2015,singh2012software,IJST70010,Rao14}. As mentioned above, this approach is
hard to test since there are very few datasets using the   pre-specified COCOMO attributes. 

Further, all that prior work needs to be revisited given the existence of recent and very prominent
methods; i.e. ATLM from TOSEM'2015~\cite{Whigham:2015} or LP4EE from TOSEM'2018~\cite{SarroTOSEM2018}.

Accordingly, this paper conducts a  more thorough investigation of   hyperparameter optimization for effort estimation. 
\bi
\item
We  use methods with no   data
feature assumptions (i.e. no COCOMO data);
\item
That
 vary
many   parameters (6,000+ combinations);
\item
That also  tests  results   on 9 different sources with data on 945 software projects; 
\item
Which  uses optimizers   representative of the  state-of-the-art 
(Differential Evolution~\cite{storn1997differential}, FLASH~\cite{nair2017flash});
\item
And which 
benchmark results 
against  prominent methods such as 
  ATLM and LP4EE.
\ei

\subsection{OIL}
\label{sec:5}

  OIL is  our  architecture for exploring hyperparameter optimization and effort estimation,
  initially, our plan
was to use standard hyperparameter  tuning for this task. Then we learned that standard machine learning toolkits like Scikit-learn~\cite{pedregosa2011scikit} did not include many of the effort estimation techniques; and (b) standard hyperparameter tuners can be  slow.  Hence, we build OIL:
 \bi
 \item
At the base {\em library layer}, we use     Scikit-learn~\cite{pedregosa2011scikit}. 
\item
Above that, OIL has a {\em utilities layer} containing all the algorithms missing in Scikit-Learn (e.g., ABEN required
numerous additions at the utilities layer). 
\item
Higher up, OIL's {\em modelling layer} uses an XML-based domain-specific language to specify a {\em feature} map of predicting model options.
These feature models are single-parent and-or graphs with (optional) cross-tree constraints showing what options require or exclude other options.
A graphical representation of  the feature model used in this paper is shown in \fig{featuretree}.
\item
Finally, at top-most {\em optimizer layer}, there is some optimizer that  makes decisions across the {\em feature} map. An automatic {\em mapper} facility then links those decisions
down to the lower layers to run the selected algorithms.  
\ei

\subsection{Optimizers}
\label{sec:optim}

Once OIL's layers were  built, it was simple  to ``pop the top'' and replace the top
layer with another optimizer.
Nair et al.~\cite{nair18}   advise that for search-based SE studies, optimizers should be selecting
via the a 
  ``dumb+two+next''  rule. Here:
  \begin{itemize}
  \item
  ``Dumb'' are some baseline methods;
  \item
  
``Two'' are some  well-established optimizers;
\item
 ``Next'' are more recent methods which may not have been
applied before to this domain.
 \end{itemize}
For our ``dumb''  optimizer, we used   Random Choice (hereafter, RD). To find $N$ valid configurations, RD
selects leaves at random from \fig{featuretree}.
All these $N$ variants are executed and the best one is selected for application to the test set. 
\RED To maintain a fair comparison with other systems described below, OIL Chooses N as the same number of evaluations in other methods. 

Moving on, our ``two'' well-established optimizers are ATLM~\cite{Whigham:2015} and LP4EE~\cite{SarroTOSEM2018}. For LP4EE, we perform experiments with the open source code provided by orginal authors. For ATLM, since there is no online source code available, we carefully re-implemented the method by ourselves.

As to our ``next'' optimizers, we used Differential Evolution (hereafter, DE~\cite{storn1997differential}) and FLASH~\cite{nair2017flash}.
The premise of DE is that the best way to mutate the existing tunings is to extrapolate between current solutions. Three solutions $a, b, c$ are selected at random. For each tuning parameter $k$, at some probability $cr$, we replace the old tuning $x_k$ with $y_k$. For
booleans $y_k = \neg x_k$ and for numerics, 
\mbox{$y_k = a_k + f \times (b_k - c_k)$}
where $f$ is a parameter controlling differential weight.  
The main loop of DE runs over the population of size $np$, replacing old items with new candidates (if new candidate is better). This means that, as the loop progresses, the population is full of increasingly more valuable solutions (which, in turn,
helps   extrapolation). 
As to the control parameters of DE,  using advice from Storn and Fu et al.~\cite{storn1997differential,Fu2016TuningFS}, we set $\{\mathit{np,g,cr}\}=\{20,0.75,0.3\}$. Also,
the number of generations $\mathit{gen}$ was set to 10 to test
the effects of  a very   CPU-light    effort estimator. 

FLASH, proposed by Nair et al.~\cite{nair2017flash}, is an incremental optimizer.
Previously, it has been applied to configuration system parameters for software systems.
This paper is the first application of FLASH to effort estimation.
Formally, FLASH is a   {\em sequential model-Based optimizer}~\cite{bergstra2011algorithms} (also known
in the machine learning literature
as an {\em active learner}~\cite{das16} or, in  the statistics literature  as 
{\em optimal experimental design}~\cite{olsson2009literature}). Whatever the name, the intuition is the same:
reflects on the model built to date in order to find the next best example
to evaluate. To tune a data miner, FLASH explores $N$ possible tunings as follows:
\begin{enumerate}
\item
Set the evaluation budget $b$. In order to make a fair comparison between FLASH and other methods, we used $b=200$.
 \item
Run the data miner using $n=20$ randomly selected tunings.
\item Build an {\em archive} of  $n$   examples holding pairs of  parameter settings and   their resulting performance scores
(e.g. MRE, SA, etc).
\item
Using that archive, learn a {\em surrogate}   to predicts performance. 
Following the methods of Nair et al.~\cite{nair2017flash}, we used   CART~\cite{brieman84}
for that surrogate.
\item Use the surrogate to guess  $M$ performance scores where
$M<N$ and $M \gg n$ parameter settings. Note that this step is very
fast since it all that is required is to run $M$ vectors down some very
small CART trees.
\item Using some {\em selection  function}, select  the most ``interesting'' setting. After Nair et al.~\cite{nair2017flash} 
we returned the setting with the nest prediction (i.e. find the most promising possibility).
\item Collect performance scores by evaluating    ``interesting'' using
the data miners (i.e. check the most troubling
possibility). Set $b=b-1$.
\item  Add  ``interesting'' to the archive. If  $b>0$, goto step 4. Else, halt.
\end{enumerate}
In summary,  given what we already know about the tunings (represented in a CART tree),
FLASH finds the potentially best   thing (in Step6); then  checks that  thing (in Step7); 
then   updates the model with the results of that check.




\BLACK

\section{Empirical Study} \label{sect:study} 

\subsection{Data}

\begin{wraptable}{r}{2.3in}
\small
\caption{Data used in this study. For details on the features, see Table~\ref{table:dataset}.}\label{table:dataset_c}
\centering
\begin{tabular}{r|rr}
 	&Projects&	Features\\\hline
kemerer	&15&	6\\
albrecht&	24&	7\\
isbsg10&    37& 11\\
finnish	&38&	7\\
miyazaki&	48&	7\\
maxwell&	62	&25\\
desharnais&	77&	6\\
kitchenham& 145&    6\\
china&  499&    16\\\hline
total & 945
\end{tabular} 
\end{wraptable}

\newcommand{\IT}[1]{\textcolor{red}{{\em #1}}}
\begin{table*}[t!]
\caption{Descriptive Statistics of the Datasets. Terms in \IT{red} are removed from this study, for reasons discussed in the text.}\label{table:dataset}
\renewcommand{\baselinestretch}{0.75} 
\resizebox{1\textwidth}{!}{
\centering
\begin{tabular}{cc}
\scriptsize
\begin{tabular}{|c|l|rrrr|}
    \hline
      & feature & min  & max & mean & std\\
   \hline


\multirow{7}{*}{\begin{sideways}kemerer\end{sideways}}
& \IT{Duration} & 5 & 31 & 14.3 & 7.5\\
& \IT{KSLOC} & 39 & 450 & 186.6 & 136.8\\
& AdjFP & 100 & 2307 & 999.1 & 589.6\\
& \IT{RAWFP} & 97 & 2284 & 993.9 & 597.4\\
& Effort & 23 & 1107 & 219.2 & 263.1\\
\hline
\multirow{8}{*}{\begin{sideways}albrecht\end{sideways}}
& Input & 7 & 193 & 40.2 & 36.9\\
& Output & 12 & 150 & 47.2 & 35.2\\
& Inquiry & 0 & 75 & 16.9 & 19.3\\
& File & 3 & 60 & 17.4 & 15.5\\
& \IT{FPAdj} & 1 & 1 & 1.0 & 0.1\\
& \IT{RawFPs} & 190 & 1902 & 638.5 & 452.7\\
& \IT{AdjFP} & 199 & 1902 & 647.6 & 488.0\\
& Effort & 0 & 105 & 21.9 & 28.4\\
\hline
\multirow{12}{*}{\begin{sideways}isbsg10\end{sideways}}
& UFP & 1 & 2 & 1.2 & 0.4\\
& IS & 1 & 10 & 3.2 & 3.0\\
& DP & 1 & 5 & 2.6 & 1.1\\
& LT & 1 & 3 & 1.6 & 0.8\\
& PPL & 1 & 14 & 5.1 & 4.1\\
& CA & 1 & 2 & 1.1 & 0.3\\
& FS & 44 & 1371 & 343.8 & 304.2\\
& RS & 1 & 4 & 1.7 & 0.9\\
& FPS & 1 & 5 & 3.5 & 0.7\\
& Effort & 87 & 14453 & 2959 & 3518\\
\hline
\multirow{8}{*}{\begin{sideways}finnish\end{sideways}}
& hw & 1 & 3 & 1.3 & 0.6\\
& at & 1 & 5 & 2.2 & 1.5\\
& FP & 65 & 1814 & 763.6 & 510.8\\
& co & 2 & 10 & 6.3 & 2.7\\
& \IT{prod} & 1 & 29 & 10.1 & 7.1\\
& \IT{lnsize} & 4 & 8 & 6.4 & 0.8\\
& \IT{lneff} & 6 & 10 & 8.4 & 1.2\\
& Effort & 460 & 26670 & 7678 & 7135\\
\hline
\multirow{19}{*}{\begin{sideways}china\end{sideways}}
& \IT{AFP} & 9 & 17518 & 486.9 & 1059\\
& Input & 0 & 9404 & 167.1 & 486.3\\
& Output & 0 & 2455 & 113.6 & 221.3\\
& Enquiry & 0 & 952 & 61.6 & 105.4\\
& File & 0 & 2955 & 91.2 & 210.3\\
& Interface & 0 & 1572 & 24.2 & 85.0\\
& \IT{Added} & 0 & 13580 & 360.4 & 829.8\\
& \IT{changed} & 0 & 5193 & 85.1 & 290.9\\
& \IT{Deleted} & 0 & 2657 & 12.4 & 124.2\\
& \IT{PDR\_A} & 0 & 84 & 11.8 & 12.1\\
& \IT{PDR\_U} & 0 & 97 & 12.1 & 12.8\\
& \IT{NPDR\_A} & 0 & 101 & 13.3 & 14.0\\
& \IT{NPDU\_U} & 0 & 108 & 13.6 & 14.8\\
& Resource & 1 & 4 & 1.5 & 0.8\\
& \IT{Dev.Type} & 0 & 0 & 0.0 & 0.0\\
& \IT{Duration} & 1 & 84 & 8.7 & 7.3\\
& Effort & 26 & 54620 & 3921 & 6481\\
\hline

\end{tabular} 


~

\scriptsize
\begin{tabular}{|c|l|rrrr|}
    \hline
      & feature
    & min  & max & mean & std\\
   \hline

\multirow{8}{*}{\begin{sideways}miyazaki\end{sideways}}
& \IT{KLOC} & 7 & 390 & 63.4 & 71.9\\
& SCRN & 0 & 150 & 28.4 & 30.4\\
& FORM & 0 & 76 & 20.9 & 18.1\\
& FILE & 2 & 100 & 27.7 & 20.4\\
& ESCRN & 0 & 2113 & 473.0 & 514.3\\
& EFORM & 0 & 1566 & 447.1 & 389.6\\
& EFILE & 57 & 3800 & 936.6 & 709.4\\
& Effort & 6 & 340 & 55.6 & 60.1\\
\hline
\multirow{26}{*}{\begin{sideways}maxwell\end{sideways}}
& App & 1 & 5 & 2.4 & 1.0\\
& Har & 1 & 5 & 2.6 & 1.0\\
& Dba & 0 & 4 & 1.0 & 0.4\\
& Ifc & 1 & 2 & 1.9 & 0.2\\
& Source & 1 & 2 & 1.9 & 0.3\\
& Telon. & 0 & 1 & 0.2 & 0.4\\
& Nlan & 1 & 4 & 2.5 & 1.0\\
& T01 & 1 & 5 & 3.0 & 1.0\\
& T02 & 1 & 5 & 3.0 & 0.7\\
& T03 & 2 & 5 & 3.0 & 0.9\\
& T04 & 2 & 5 & 3.2 & 0.7\\
& T05 & 1 & 5 & 3.0 & 0.7\\
& T06 & 1 & 4 & 2.9 & 0.7\\
& T07 & 1 & 5 & 3.2 & 0.9\\
& T08 & 2 & 5 & 3.8 & 1.0\\
& T09 & 2 & 5 & 4.1 & 0.7\\
& T10 & 2 & 5 & 3.6 & 0.9\\
& T11 & 2 & 5 & 3.4 & 1.0\\
& T12 & 2 & 5 & 3.8 & 0.7\\
& T13 & 1 & 5 & 3.1 & 1.0\\
& T14 & 1 & 5 & 3.3 & 1.0\\
& T15 & 1 & 5 & 3.3 & 0.7\\
& \IT{Dura.} & 4 & 54 & 17.2 & 10.7\\
& Size & 48 & 3643 & 673.3 & 784.1\\
& \IT{Time} & 1 & 9 & 5.6 & 2.1\\
& Effort & 583 & 63694 & 8223 & 10500\\
\hline
\multirow{7}{*}{\begin{sideways}desharnais\end{sideways}}
& TeamExp & 0 & 4 & 2.3 & 1.3\\
& MngExp & 0 & 7 & 2.6 & 1.5\\
& \IT{Length} & 1 & 36 & 11.3 & 6.8\\
& Trans.s & 9 & 886 & 177.5 & 146.1\\
& Entities & 7 & 387 & 120.5 & 86.1\\
& AdjPts & 73 & 1127 & 298.0 & 182.3\\
& Effort & 546 & 23940 & 4834 & 4188\\
\hline
\multirow{7}{*}{\begin{sideways}kitchenham\end{sideways}}
& code & 1 & 6 & 2.1 & 0.9\\
& type & 0 & 6 & 2.4 & 0.9\\
& \IT{duration} & 37 & 946 & 206.4 & 134.1\\
& fun\_pts & 15 & 18137 & 527.7 & 1522\\
& \IT{estimate} & 121 & 79870 & 2856 & 6789\\
& \IT{esti\_mtd} & 1 & 5 & 2.5 & 0.9\\
& Effort & 219 & 113930 & 3113 & 9598\\
\hline

\end{tabular} 
\end{tabular}
}
\end{table*}

To assess OIL, we applied it to the 945 projects
seen in nine    datasets from the SEACRAFT repository\footnote{http://tiny.cc/seacraft}; see Table~\ref{table:dataset_c} and Table~\ref{table:dataset}. 
This data was selected since it has been  widely  used in previous estimation research.
Also, it  is quite diverse since it differs in
  number of observations (from 15 to 499 projects);
  geographical locations (software projects coming from Canada, China, Finland);
 technical characteristics (software projects developed in different programming languages and for different application domains, ranging from telecommunications to commercial information systems);
and 
  number and type of  features (from 6 to 25  features, including a variety of   features describing the software projects, such as number of developers involved in the project and their experience, technologies used, size in terms of Function Points, etc.);

\RED Note that some features of the original datasets are not used in our experiment because they are (1) naturally irrelevant to their effort values (e.g., ID, Syear), (2) unavailable at the prediction phase (e.g., duration, LOC), (3) highly correlated or overlap to each other (e.g., raw function point and adjusted function points). A data cleaning process is applied to solve this issue. 
Those removed features are shown as italic in Table~\ref{table:dataset}. 

\BLACK
\subsection{Cross-Validation}

Each datasets was treated in a variety of  ways. Each {\em treatment} is an {\em M*N-way} cross-validation test of some learners or some learners and optimizers. That is, $M$ times,  shuffle the data randomly (using a different random number seed)
then divide the data into $N$ bins.
For $i   \in N$, bin $i$ is used to test a model
build from the other bins.
Following the advice
of Nair et al.~\cite{nair18}, we  use $N=3$  bins
for our effort datasets.   

As a procedural detail, first we divided the data and then we applied the treatments. That is, all treatments saw the same training and test data.

\subsection{Scoring Metrics}

\begin{table}
\caption{Performance scores: MRE and SA}\label{samre}
\begin{tabular}{|p{.95\linewidth}|}\hline
{\bf MRE:}
MRE is defined in terms of 
AR,  the magnitude of the absolute residual. This is  computed from the difference between predicted and actual effort values:
\[
\mathit{AR} = |\mathit{actual}_i - \mathit{predicted}_i|
\] 
MRE is the magnitude of the relative error calculated by expressing AR as a ratio of   actual effort:
\[
\mathit{MRE} = \frac{|\mathit{actual}_i - \mathit{predicted}_i|}{\mathit{actual}_i}
\]
MRE has been criticized~\cite{foss2003simulation,kitchenham2001accuracy,korte2008confidence,port2008comparative,shepperd2000building,stensrud2003further} as being biased towards error underestimations. Nevertheless, we use it here
since  there exists known baselines for human performance in effort estimation, expressed in terms of MRE~\cite{Jorgensen03}. The same can not be said for SA.
\\\hline
{\bf SA:}
Because of issues with MRE, some researchers prefer the 
use of other (more standardized) measures, such as  Standardized Accuracy (SA)~\cite{langdon2016exact,shepperd2012evaluating}.
SA is defined in terms of 
\[
\mathit{MAE}=\frac{1}{N}\sum_{i=1}^n|\mathit{RE}_i-\mathit{EE}_i|
\]
where $N$ is the number of projects used for evaluating the performance, and $\mathit{RE}_i$ and $\mathit{EE}_i$ are the actual and estimated effort, respectively, for the project $i$. 
SA uses MAE as follows:
\[
\mathit{SA} = (1-\frac{\mathit{MAE}_{P_{j}}}{\mathit{MAE}_{r_{guess}}})\times 100
\]
where $\mathit{MAE}_{P_{j}}$ is the MAE of the approach $P_j$ being evaluated and $\mathit{MAE}_{r_{\mathit{guess}}}$ is the MAE of a large number (e.g., 1000 runs) of random guesses. 
Over many runs,  $\mathit{MAE}_{r_{\mathit{guess}}}$ will converge on simply using the sample mean~\cite{shepperd2012evaluating}. That is, SA represents how much better $P_j$ is than random guessing. Values near zero means that the prediction model $P_j$ is practically useless, performing little better than  random guesses~\cite{shepperd2012evaluating}. \\\hline
\end{tabular}
\end{table}

\begin{table}
\caption{Explanation of Scott-Knott test.}\label{tbl:stats}

\begin{tabular}{|p{0.96\linewidth}|}\hline
\small

This study ranks methods using the Scott-Knott
procedure recommended by Mittas \& Angelis in their 2013
IEEE TSE paper~\cite{Mittas13}.  This method
sorts a list of $l$ treatments with $\mathit{ls}$ measurements by their median
score. It then
splits $l$ into sub-lists $m,n$ in order to maximize the expected value of
 differences  in the observed performances
before and after divisions. For example, we could sort $\mathit{ls}=4$ 
methods based on their median score,
then divide them into three sub-lists of of size $\mathit{ms},\mathit{ns} \in \{(1,3), (2,2), (3,1)\}$.
Scott-Knott would declare one of these divisions
to be ``best'' as follows.
For lists $l,m,n$ of size $\mathit{ls},\mathit{ms},\mathit{ns}$ where $l=m\cup n$, the ``best'' division maximizes $E(\Delta)$; i.e.
the difference in the expected mean value
before and after the spit: 
 \[E(\Delta)=\frac{ms}{ls}abs(m.\mu - l.\mu)^2 + \frac{ns}{ls}abs(n.\mu - l.\mu)^2\]
Scott-Knott then checks if that
``best'' division is actually useful.
To implement that check, Scott-Knott would
apply some statistical hypothesis test $H$ to check
if $m,n$ are significantly different. If so, Scott-Knott then recurses on each half of the ``best'' division.

For a more specific example, consider the results
from $l=5$ treatments:

{\scriptsize \begin{verbatim}
        rx1 = [0.34, 0.49, 0.51, 0.6]
        rx2 = [0.6,  0.7,  0.8,  0.9]
        rx3 = [0.15, 0.25, 0.4,  0.35]
        rx4=  [0.6,  0.7,  0.8,  0.9]
        rx5=  [0.1,  0.2,  0.3,  0.4]
\end{verbatim}}
\noindent
After sorting and division, Scott-Knott declares:
\bi
\item Ranked \#1 is rx5 with median= 0.25
\item Ranked \#1 is rx3 with median= 0.3
\item Ranked \#2 is rx1 with median= 0.5
\item Ranked \#3 is rx2 with median= 0.75
\item Ranked \#3 is rx4 with median= 0.75
\ei
Note that Scott-Knott found  little
difference between rx5 and rx3. Hence,
they have the same rank, even though their medians differ.

Scott-Knott is prefered to, say, an 
 all-pairs hypothesis test of all methods; e.g. six treatments
can be compared \mbox{$(6^2-6)/2=15$} ways. 
A 95\% confidence test run for each comparison has  a very low total confidence: 
\mbox{$0.95^{15} = 46$}\%.
To avoid an all-pairs comparison, Scott-Knott only calls on hypothesis
tests {\em after} it has found splits that maximize the performance differences.
 
For this study, our hypothesis test $H$ was a
conjunction of the A12 effect size test of  and
non-parametric bootstrap sampling; i.e. our
Scott-Knott divided the data if {\em both}
bootstrapping and an effect size test agreed that
the division was statistically significant (95\%
confidence) and not a ``small'' effect ($A12 \ge
0.6$).

For a justification of the use of non-parametric
bootstrapping, see Efron \&
Tibshirani~\cite[p220-223]{efron93}.
For a justification of the use of effect size tests
see Shepperd \& MacDonell~\cite{shepperd2012evaluating}; Kampenes~\cite{kampenes2007}; and
Kocaguneli et al.~\cite{Keung2013}. These researchers
warn that even if an
hypothesis test declares two populations to be
``significantly'' different, then that result is
misleading if the ``effect size'' is very small.
Hence, to assess 
the performance differences 
we first must rule out small effects.
Vargha and Delaney's
non-parametric 
A12 effect size test 
explores
two lists $M$ and $N$ of size $m$ and $n$:

\[
A12 = \left(\sum_{x\in M, y \in N} 
\begin{cases} 
1   & \mathit{if}\; x > y\\
0.5 & \mathit{if}\; x == y
\end{cases}\right) / (mn)
\]

This expression computes the probability that numbers in one sample are bigger than in another.
This test was recently 
endorsed by Arcuri and Briand
at ICSE'11~\cite{arcuri2011}.

\\\hline
\end{tabular}
\end{table}

The results from each test set are evaluated in terms of two scoring metrics:  magnitude of the relative error (MRE)~\cite{Conte:1986:SEM:6176} and Standardized Accuracy (SA). These scoring metrics  are defined in Table~\ref{samre}.
We use these since there are advocates for each in the literature.
For example, Shepperd and MacDonell argue convincingly for the use of
 SA~\cite{shepperd2012evaluating} (as well as for the use of effect size tests in effort estimation).
 Also in 2016, Sarro et al.\footnote{http://tiny.cc/sarro16gecco} used MRE  to argue their estimators were competitive with human estimates
 (which Molokken et al.~\cite{molokken2003review} says lies within 30\% and 40\% of the true value).

Note that for these evaluation measures:
\bi
\item MRE values: {\em smaller} are {\em better}
\item SA values: {\em larger} are {\em better}
\ei

From the cross-vals,
we  report the {\em median} (termed {\em med})
which is the 50th percentile of the   test scores seen in the {\em M*N results}.
Also reported are the  {\em inter-quartile range} (termed  {\em IQR}) which is the (75-25)th percentile.
The IQR is a  non-parametric
description of the   variability about the median value.  

For each datasets, the results from a {\em M*N-way} are sorted by their {\em  median} value, then {\em ranked} using the Scott-Knott test
recommended for ranking effort estimation experiments by Mittas et al. in TSE'13~\cite{Mittas13}. 
For full details on Scott-Knott test, see Table~\ref{tbl:stats}. In summary, Scott-Knott is a top-down bi-clustering
method that recursively divides sorted treatments. Division stops when there is only one treatment left or when a division of numerous treatments generates 
splits that are statistically {\em indistinguishable}. 
To judge when two sets of treatments are indistinguishable, we use a conjunction of {\em both}  a 95\% bootstrap significance test~\cite{efron93} {\em and}
a A12 test for a non-small effect size difference in the distributions~\cite{MenziesNeg:2017}. These tests were used since their non-parametric nature avoids issues with non-Gaussian
distributions. 

Table~\ref{eg} shows an example of the report generated by our Scott-Knott procedure.
Note that when multiple treatments tie for {\em Rank=1}, then we use the treatment's
runtimes to break the tie. Specifically, for all treatments in {\em Rank=1}, we mark the faster ones as \colorbox{blue!10}{{\em Rank=1}}.

\begin{table}[!h]
\centering

 \caption{Example of Scott-Knott results.
 SA scores seen in  the albrecht dataset.
 sorted by their median value. 
 Here, {\em larger} values are {\em better}.
  {\bf Med} is the 50th percentile and {\bf IQR} is the {\em inter-quartile range}; i.e., 75th-25th percentile. 
    Lines with a dot in the middle  shows   median values with the IQR.   
  For the  {\bf Ranks},  {\em smaller} values are  {\em better}.
   Ranks are computed via the Scott-Knot procedure from  TSE'2013~\cite{Mittas13}.
    Rows with the same ranks
    are statistically indistinguishable. 
    Rows shown in 
  \colorbox{blue!10}{color} denotes rows of fastest best-ranked treatments.}\label{eg}
{\scriptsize   
\begin{tabular}{llc@{~~~}c@{~~~}c} 
&  &\multicolumn{2}{c}{\textbf{Standardized Accuracy}} & \\ 
  {\textbf{Rank}}& \textbf{Method} & \textbf{Med} & \textbf{IQR} & \\\hline
  
    \rowcolor{blue!10}  1 &      CART\_FLASH &    65 &  18 & \quart{54}{18}{65}{100} \\
    \rowcolor{blue!10}  1 &      CART\_DE &    59 &  19 & \quart{48}{19}{59}{100} \\
    2 &      ABEN\_DE &    52 &  23 & \quart{42}{23}{52}{100} \\
    2 &      ABE0 &    51 &  20 & \quart{43}{20}{51}{100} \\
    2 &      RF &    49 &  29 & \quart{34}{29}{49}{100} \\
    2 &      LP4EE &    47 &  25 & \quart{32}{25}{47}{100} \\
    3 &      CART &    41 &  31 & \quart{23}{31}{41}{100} \\
    3 &      ABEN\_RD &    37 &  33 & \quart{26}{33}{37}{100} \\
    3 &      ATLM &    34 &  13 & \quart{31}{13}{34}{100} \\
    3 &      CART\_RD &    32 &  27 & \quart{21}{27}{32}{100} \\
    3 &      SVR &    30 &  18 & \quart{24}{18}{30}{100} \\
    \hline   

 \end{tabular}}

\end{table}

\subsection{Terminology for Optimizers}

Some treatments are named ``X\_Y'' which  denote learner ``X'' tuned by optimizer ``Y''.
In the following: 

{\small \begin{eqnarray} 
X &\in &\{\mathit{CART},\mathit{ABE}\}\nonumber\\
Y &\in &\{\mathit{DE},\mathit{RD},\mathit{FLASH}\}\nonumber
\end{eqnarray}}
Note that we do not tune ATLM and LP4EE since they were designed to be used ``off-the-shelf''.  Whigham et al.~\cite{Whigham:2015} declare that one of ATLM's most important features is that if does not need tuning.
\RED We also do not tune SVR and RF since we treat them as baseline algorithm-based methods in our benchmarks (i.e. use default settings in scikit-learn for these algorithms).
\BLACK

\section{Results}

\newcommand{\PP}{$< \! \! 1$}
\newcommand{\PM}{$< \! \! 30$}
\begin{table}
\caption{Average runtime (in minutes), for one-way out of an N*M cross-validation
experiment. cross-validation (minutes). Executing on a 2GHz processor, 
with 8GB RAM,  running Windows 10.
Note that LP4EE and ATLM have no tuning results since the authors of these methods stress that it is advantageous to use their baseline methods, without any tuning. Last column reports totals for each dataset.  }\label{tbl:runtime}
\vspace{3mm}
\small
\begin{center}

\begin{tabular}{|r@{~}|r@{~}r@{~}r@{~}r@{~}r@{~}r@{~}r@{~}r@{~}r@{~}|r@{~}r|r|}\cline{2-12}
\multicolumn{1}{c|}{~}&\multicolumn{9}{c|}{faster}&\multicolumn{2}{c|}{slower}\\\cline{2-12}
\multicolumn{1}{c|}{~}&	\begin{turn}{75}ABE0\end{turn}	
&	\begin{turn}{75}CART\end{turn}	
&	\begin{turn}{75}ATLM\end{turn}
&	\begin{turn}{75}LP4EE\end{turn}	
&	\begin{turn}{75}SVR\end{turn}	
&	\begin{turn}{75}RF\end{turn}
&   \begin{turn}{75}CART\_RD\end{turn}	
&	\begin{turn}{75}CART\_DE\end{turn}	
&	\begin{turn}{75}CART\_FLASH\end{turn}	
&	\begin{turn}{75}ABEN\_RD\end{turn}	
&	\begin{turn}{75}ABEN\_DE\end{turn}
&   \multicolumn{1}{r}{	 total } \\\hline
kemerer	&   \PP	&\PP	&\PP	&	\PP&\PP&\PP&\PP&\PP&\PP  &4  &5   &13\\
albrecht&	\PP	&\PP	&\PP	&	\PP&\PP&\PP&\PP&\PP&\PP  &4  &6   &15\\
finnish&	\PP	&\PP	&\PP	&	\PP&\PP&\PP&\PP&\PP&\PP  &5  &6   &18\\
miyazaki&	\PP	&\PP	&\PP	&	\PP&\PP&\PP&\PP&\PP&\PP  &6  &8   &21\\
desharnais&	\PP	&\PP	&\PP	&	\PP&\PP&\PP&\PP&\PP&\PP  &9  &11   &24\\
isbsg10&	\PP	&\PP	&\PP	&	\PP&\PP&\PP&\PP&\PP&\PP  &7  &10   &23\\
maxwell	&   \PP	&\PP	&\PP	&	\PP&\PP&\PP&\PP&\PP&\PP  &12  &16  &34\\
kitchenham&	\PP	&\PP	&\PP	&	\PP&\PP&\PP&\PP&\PP&\PP  &16  &17  &37\\
china&   	\PP	&\PP	&\PP	&	\PP&\PP&\PP&\PP&\PP&\PP  &23  &26  &54\\\hline													
total	&3 &4 &4 &4 &5 &7 &8 &9 &8 &86 &105\\\cline{1-12}
\end{tabular}
\end{center}
\end{table}

\begin{table}
 \caption{
\%  {\bf MRE} results
from our cross-validation studies. {\em Smaller} values are {\em better}.
Same format as Table~\ref{eg}.
The gray rows show the \colorbox{blue!10}{{\em Rank=1} results} recommended
for each data set.
The phrase ``\ofr'' denotes results
that are so bad that they fall outside of the 0\%..100\% range shown here.
}\label{table_mre}
\centering
\renewcommand{\baselinestretch}{0.4} 
\resizebox{0.7\textwidth}{!}{
{
\scriptsize
\noindent
\begin{tabular}{p{1.5cm}llrc}
{\textbf{Rank}} & \textbf{Method} & \textbf{Med.} & \textbf{IQR}\\
 \hline

    \nm{albrecht}\\
    \rowcolor{blue!10} 1 & CART\_FLASH &    32 &  12 & \quart{26}{12}{32}{100} \\
    \rowcolor{blue!10}1 &      CART\_DE &    33 &  14 & \quart{27}{14}{33}{100} \\
    2 &      ABEN\_DE &    42 &  15 & \quart{36}{15}{42}{100} \\
    2 &      LP4EE &    44 &  13 & \quart{39}{13}{44}{100} \\
    2 &      ABE0 &    45 &  16 & \quart{37}{16}{45}{100} \\
    2 &      RF &    46 &  24 & \quart{35}{24}{46}{100} \\
    2 &      ABEN\_RD &    48 &  21 & \quart{37}{21}{48}{100} \\
    3 &      CART &    53 &  22 & \quart{43}{22}{53}{100} \\
    3 &      CART\_RD &    54 &  20 & \quart{44}{20}{54}{100} \\
    3 &      SVR &    56 &  19 & \quart{45}{19}{56}{100} \\
    4 &      ATLM &    140 &  91 & \ofr \\
    \hline

  \nm{china}\\
  \rowcolor{blue!10}  1 &      LP4EE &    45 &  5 & \quart{42}{5}{45}{100} \\
  \rowcolor{blue!10}1 &      ATLM &    48 &  6 & \quart{45}{6}{48}{100} \\
2 &      CART\_FLASH &    61 &  5 & \quart{59}{5}{61}{100} \\
2 &      ABEN\_DE &    62 &  6 & \quart{58}{6}{62}{100} \\
3 &      ABE0 &    64 &  5 & \quart{61}{5}{64}{100} \\
3 &      CART\_DE &    64 &  6 & \quart{61}{6}{64}{100} \\
4 &      ABEN\_RD &    68 &  7 & \quart{64}{7}{68}{100} \\
4 &      RF &    69 &  8 & \quart{65}{8}{69}{100} \\
5 &      SVR &    71 &  7 & \quart{68}{7}{71}{100} \\
5 &      CART &    71 &  5 & \quart{68}{5}{71}{100} \\
5 &      CART\_RD &    72 &  6 & \quart{69}{6}{72}{100} \\
    \hline
 
  \nm{desharnais}\\
  \rowcolor{blue!10}1 &      CART\_DE &    35 &  11 & \quart{30}{11}{35}{100} \\
  \rowcolor{blue!10} 1 &      CART\_FLASH &    35 &  11 & \quart{30}{11}{35}{100} \\
  \rowcolor{blue!10} 1 &      LP4EE &    38 &  13 & \quart{34}{13}{38}{100} \\
2 &      RF &    46 &  12 & \quart{41}{12}{46}{100} \\
2 &      ABEN\_DE &    47 &  14 & \quart{43}{14}{47}{100} \\
2 &      SVR &    48 &  12 & \quart{42}{12}{48}{100} \\
2 &      CART\_RD &    48 &  13 & \quart{44}{13}{48}{100} \\
2 &      ABEN\_RD &    49 &  16 & \quart{39}{16}{49}{100} \\
2 &      CART &    49 &  14 & \quart{43}{14}{49}{100} \\
2 &      ABE0 &    50 &  15 & \quart{39}{15}{50}{100} \\
2 &      ATLM &    54 &  17 & \quart{37}{17}{54}{100} \\
    \hline
    
    \nm{finnish}\\
    \rowcolor{blue!10}1 &      CART\_FLASH &    42 &  17 & \quart{35}{17}{42}{100} \\
2 &      CART\_DE &    48 &  16 & \quart{40}{16}{48}{100} \\
3 &      CART &    57 &  21 & \quart{53}{21}{57}{100} \\
3 &      RF &    57 &  26 & \quart{45}{26}{57}{100} \\
3 &      CART\_RD &    58 &  22 & \quart{46}{22}{58}{100} \\
4 &      ABEN\_DE &    62 &  37 & \quart{55}{37}{62}{100} \\
4 &      LP4EE &    63 &  33 & \quart{57}{33}{63}{100} \\
4 &      ABE0 &    64 &  48 & \quart{55}{48}{64}{100} \\
4 &      ABEN\_RD &    64 &  42 & \quart{51}{42}{64}{100} \\
4 &      SVR &    74 &  13 & \quart{67}{13}{74}{100} \\
5 &      ATLM &    87 &  72 & \quart{49}{56}{87}{100} \\
    \hline
    
     \nm{isbsg10}\\
     \rowcolor{blue!10}1 &      CART\_DE &    59 &  20 & \quart{49}{20}{59}{100} \\
     \rowcolor{blue!10}1 &      CART\_FLASH &    62 &  19 & \quart{51}{19}{62}{100} \\
2 &      SVR &    72 &  17 & \quart{61}{17}{72}{100} \\
2 &      CART\_RD &    73 &  24 & \quart{62}{24}{73}{100} \\
2 &      ABE0 &    73 &  60 & \quart{61}{44}{73}{100} \\
2 &      CART &    74 &  21 & \quart{66}{21}{74}{100} \\
2 &      ABEN\_DE &    74 &  25 & \quart{60}{25}{74}{100} \\
2 &      LP4EE &    75 &  23 & \quart{64}{23}{75}{100} \\
2 &      ABEN\_RD &    76 &  35 & \quart{51}{35}{76}{100} \\
2 &      RF &    78 &  58 & \quart{69}{36}{78}{100} \\
3 &      ATLM &    127 &  124 & \ofr \\
    \hline 
    
     \nm{kemerer}\\
     \rowcolor{blue!10} 1 &      CART\_DE &    32 &  24 & \quart{21}{24}{32}{100} \\
     \rowcolor{blue!10}1 &      CART\_FLASH &    37 &  27 & \quart{26}{27}{37}{100} \\
2 &      RF &    50 &  39 & \quart{35}{39}{50}{100} \\
2 &      ABEN\_DE &    54 &  22 & \quart{35}{22}{54}{100} \\
2 &      LP4EE &    54 &  23 & \quart{36}{23}{54}{100} \\
2 &      CART\_RD &    55 &  25 & \quart{37}{25}{55}{100} \\
2 &      CART &    55 &  27 & \quart{37}{27}{55}{100} \\
3 &      ABE0 &    56 &  33 & \quart{46}{33}{56}{100} \\
3 &      SVR &    59 &  14 & \quart{56}{14}{59}{100} \\
3 &      ABEN\_RD &    60 &  17 & \quart{54}{17}{60}{100} \\
4 &      ATLM &    76 &  56 & \quart{45}{56}{76}{100} \\
    \hline
 
    \nm{kitchenham}\\
    \rowcolor{blue!10}1 &      CART\_FLASH &    34 &  6 & \quart{32}{6}{34}{100} \\
    \rowcolor{blue!10} 1 &      ABEN\_DE &    35 &  8 & \quart{32}{8}{35}{100} \\
2 &      LP4EE &    38 &  6 & \quart{34}{6}{38}{100} \\
2 &      CART\_DE &    38 &  7 & \quart{35}{7}{38}{100} \\
2 &      ABE0 &    39 &  9 & \quart{36}{9}{39}{100} \\
3 &      ABEN\_RD &    42 &  10 & \quart{36}{10}{42}{100} \\
3 &      RF &    43 &  8 & \quart{39}{8}{43}{100} \\
4 &      CART &    49 &  11 & \quart{43}{11}{49}{100} \\
5 &      CART\_RD &    57 &  12 & \quart{51}{12}{57}{100} \\
5 &      SVR &    60 &  8 & \quart{55}{8}{60}{100} \\
6 &      ATLM &    106 &  108 & \ofr \\
    \hline 
    
    \nm{maxwell}\\
    \rowcolor{blue!10}1 &      CART\_FLASH &    36 &  10 & \quart{30}{10}{36}{100} \\
    \rowcolor{blue!10}1 &      CART\_DE &    38 &  8 & \quart{34}{8}{38}{100} \\
2 &      RF &    51 &  15 & \quart{42}{15}{51}{100} \\
2 &      LP4EE &    51 &  16 & \quart{44}{16}{51}{100} \\
2 &      CART &    52 &  10 & \quart{46}{10}{52}{100} \\
2 &      CART\_RD &    53 &  11 & \quart{45}{11}{53}{100} \\
2 &      ABEN\_DE &    53 &  16 & \quart{43}{16}{53}{100} \\
3 &      SVR &    56 &  13 & \quart{50}{13}{56}{100} \\
3 &      ABEN\_RD &    56 &  13 & \quart{49}{13}{56}{100} \\
3 &      ABE0 &    56 &  14 & \quart{48}{14}{56}{100} \\
4 &      ATLM &    282 &  221 & \ofr \\
    \hline 
    
    \nm{miyazaki}\\
    \rowcolor{blue!10} 1 &      CART\_DE &    32 &  11 & \quart{27}{11}{32}{100} \\
    \rowcolor{blue!10} 1 &      CART\_FLASH &    32 &  11 & \quart{27}{11}{32}{100} \\
    \rowcolor{blue!10} 1 &      LP4EE &    33 &  10 & \quart{28}{10}{33}{100} \\
2 &      SVR &    37 &  10 & \quart{33}{10}{37}{100} \\
2 &      ATLM &    37 &  32 & \quart{25}{32}{37}{100} \\
2 &      ABEN\_DE &    39 &  16 & \quart{30}{16}{39}{100} \\
3 &      RF &    46 &  24 & \quart{35}{24}{46}{100} \\
3 &      ABE0 &    47 &  12 & \quart{39}{12}{47}{100} \\
3 &      CART &    47 &  16 & \quart{39}{16}{47}{100} \\
3 &      ABEN\_RD &    47 &  15 & \quart{39}{15}{47}{100} \\
3 &      CART\_RD &    48 &  14 & \quart{40}{14}{48}{100} \\
    \hline

  \end{tabular}
}}
\end{table}

\begin{table}
 \caption{
  \% {\bf SA} results
from our cross-validation studies. {\em Larger} values are {\em better}. Same format as Table~\ref{eg}.
The gray rows show the \colorbox{blue!10}{{\em Rank=1} results} recommended
for each data set.
The phrase ``\ofr'' denotes results
that are so bad that they fall outside of the 0\%..100\% range shown here.
}\label{table_sa}
\centering
\renewcommand{\baselinestretch}{0.4} 
\resizebox{0.7\textwidth}{!}{
{
\scriptsize
\noindent \begin{tabular}{p{1.5cm}llrc}
{\textbf{Rank}}& \textbf{Method} & \textbf{Med.} & \textbf{IQR}\\
 \hline
  
  \nm{albrecht}\\
  \rowcolor{blue!10} 1 &      CART\_FLASH &    65 &  18 & \quart{54}{18}{65}{100} \\
  \rowcolor{blue!10}1 &      CART\_DE &    59 &  19 & \quart{48}{19}{59}{100} \\
2 &      ABEN\_DE &    52 &  23 & \quart{42}{23}{52}{100} \\
2 &      ABE0 &    51 &  20 & \quart{43}{20}{51}{100} \\
2 &      RF &    49 &  29 & \quart{34}{29}{49}{100} \\
2 &      LP4EE &    47 &  25 & \quart{32}{25}{47}{100} \\
3 &      CART &    41 &  31 & \quart{23}{31}{41}{100} \\
3 &      ABEN\_RD &    37 &  33 & \quart{26}{33}{37}{100} \\
3 &      ATLM &    34 &  13 & \quart{31}{13}{34}{100} \\
3 &      CART\_RD &    32 &  27 & \quart{21}{27}{32}{100} \\
3 &      SVR &    30 &  18 & \quart{24}{18}{30}{100} \\
    \hline
 
    \nm{china}\\
    \rowcolor{blue!10}1 &      LP4EE &    32 &  9 & \quart{24}{9}{32}{100} \\
    \rowcolor{blue!10}1 &      CART\_FLASH &    30 &  8 & \quart{25}{8}{30}{100} \\
1 &      ABEN\_DE &    29 &  13 & \quart{23}{13}{29}{100} \\
\rowcolor{blue!10}  1 &      ABE0 &    28 &  7 & \quart{25}{7}{28}{100} \\
\rowcolor{blue!10}  1 &      CART\_DE &    27 &  6 & \quart{24}{6}{27}{100} \\
2 &      SVR &    21 &  3 & \quart{19}{3}{21}{100} \\
2 &      RF &    21 &  11 & \quart{15}{11}{21}{100} \\
2 &      ABEN\_RD &    20 &  9 & \quart{17}{9}{20}{100} \\
3 &      ATLM &    12 &  5 & \quart{10}{5}{12}{100} \\
3 &      CART &    12 &  13 & \quart{6}{13}{12}{100} \\
3 &      CART\_RD &    10 &  16 & \quart{7}{16}{10}{100} \\
    \hline
 
    \nm{desharnais}\\
    \rowcolor{blue!10} 1 &      CART\_FLASH &    53 &  11 & \quart{46}{11}{53}{100} \\
    \rowcolor{blue!10}1 &      CART\_DE &    53 &  11 & \quart{45}{11}{53}{100} \\
2 &      LP4EE &    48 &  12 & \quart{41}{12}{48}{100} \\
2 &      RF &    46 &  11 & \quart{39}{11}{46}{100} \\
2 &      ABEN\_DE &    46 &  14 & \quart{39}{14}{46}{100} \\
2 &      ABE0 &    44 &  12 & \quart{38}{12}{44}{100} \\
2 &      SVR &    43 &  7 & \quart{39}{7}{43}{100} \\
3 &      CART &    39 &  12 & \quart{32}{12}{39}{100} \\
3 &      ATLM &    37 &  8 & \quart{31}{8}{37}{100} \\
3 &      ABEN\_RD &    37 &  13 & \quart{28}{13}{37}{100} \\
4 &      CART\_RD &    31 &  10 & \quart{26}{10}{31}{100} \\
    \hline

  \nm{finnish}\\
  \rowcolor{blue!10} 1 &      CART\_FLASH &    54 &  13 & \quart{48}{13}{54}{100} \\
2 &      CART\_DE &    49 &  12 & \quart{43}{12}{49}{100} \\
3 &      RF &    44 &  16 & \quart{33}{16}{44}{100} \\
3 &      ABEN\_DE &    43 &  25 & \quart{31}{25}{43}{100} \\
3 &      CART &    42 &  17 & \quart{32}{17}{42}{100} \\
4 &      ATLM &    41 &  48 & \quart{4}{48}{41}{100} \\
4 &      CART\_RD &    40 &  22 & \quart{29}{22}{40}{100} \\
4 &      ABE0 &    40 &  25 & \quart{24}{25}{40}{100} \\
4 &      LP4EE &    39 &  22 & \quart{27}{22}{39}{100} \\
4 &      ABEN\_RD &    38 &  27 & \quart{26}{27}{38}{100} \\
5 &      SVR &    24 &  9 & \quart{19}{9}{24}{100} \\
    \hline
 
     \nm{isbsg10}\\
     \rowcolor{blue!10} 1 &      CART\_DE &    33 &  19 & \quart{21}{19}{33}{100} \\
     \rowcolor{blue!10}1 &      CART\_FLASH &    30 &  18 & \quart{21}{18}{30}{100} \\
     \rowcolor{blue!10}1 &      ATLM &    30 &  20 & \quart{25}{20}{30}{100} \\
2 &      ABEN\_DE &    28 &  24 & \quart{16}{24}{28}{100} \\
2 &      ABE0 &    28 &  23 & \quart{15}{23}{28}{100} \\
2 &      SVR &    25 &  11 & \quart{17}{11}{25}{100} \\
3 &      LP4EE &    22 &  23 & \quart{7}{23}{22}{100} \\
3 &      CART\_RD &    22 &  28 & \quart{6}{28}{22}{100} \\
3 &      RF &    22 &  35 & \quart{-3}{35}{22}{100} \\
3 &      ABEN\_RD &    21 &  27 & \quart{-3}{27}{21}{100} \\
3 &      CART &    20 &  34 & \quart{-2}{34}{20}{100} \\
    \hline
 
  \nm{kemerer}\\
  \rowcolor{blue!10}1 &      CART\_DE &    55 &  30 & \quart{39}{30}{55}{100} \\
2 &      CART\_FLASH &    43 &  27 & \quart{30}{27}{43}{100} \\
2 &      CART &    42 &  25 & \quart{31}{25}{42}{100} \\
2 &      RF &    41 &  29 & \quart{31}{29}{41}{100} \\
2 &      ABEN\_DE &    40 &  27 & \quart{31}{27}{40}{100} \\
2 &      LP4EE &    40 &  23 & \quart{30}{23}{40}{100} \\
2 &      ABE0 &    38 &  25 & \quart{28}{25}{38}{100} \\
2 &      CART\_RD &    36 &  27 & \quart{27}{27}{36}{100} \\
3 &      ABEN\_RD &    32 &  28 & \quart{24}{28}{32}{100} \\
3 &      ATLM &    30 &  28 & \quart{23}{28}{30}{100} \\
3 &      SVR &    28 &  24 & \quart{20}{24}{28}{100} \\
    \hline

  \nm{kitchenham}\\
  \rowcolor{blue!10} 1 &      LP4EE &    52 &  24 & \quart{35}{24}{52}{100} \\
1 &      ABEN\_DE &    51 &  25 & \quart{33}{25}{51}{100} \\
\rowcolor{blue!10} 1 &      ABE0 &    47 &  23 & \quart{31}{23}{47}{100} \\
\rowcolor{blue!10} 1 &      CART\_FLASH &    44 &  24 & \quart{30}{24}{44}{100} \\
2 &      RF &    41 &  19 & \quart{30}{19}{41}{100} \\
2 &      ABEN\_RD &    40 &  19 & \quart{32}{19}{40}{100} \\
2 &      CART\_DE &    40 &  20 & \quart{30}{20}{40}{100} \\
3 &      CART &    34 &  14 & \quart{26}{14}{34}{100} \\ 
4 &      CART\_RD &    32 &  18 & \quart{21}{18}{32}{100} \\
4 &      SVR &    32 &  17 & \quart{20}{17}{32}{100} \\
5 &      ATLM &    -3 &  37 & \quart{-8}{25}{-3}{100} \\  
    \hline
    
  \nm{maxwell}\\
  \rowcolor{blue!10} 1 &      CART\_FLASH &    55 &  7 & \quart{50}{7}{55}{100} \\
  \rowcolor{blue!10}1 &      LP4EE &    52 &  13 & \quart{47}{13}{52}{100} \\
  \rowcolor{blue!10} 1 &      CART\_DE &    51 &  16 & \quart{44}{16}{51}{100} \\
2 &      RF &    44 &  11 & \quart{38}{11}{44}{100} \\
2 &      ABEN\_DE &    43 &  12 & \quart{37}{12}{43}{100} \\
3 &      ABE0 &    39 &  10 & \quart{33}{10}{39}{100} \\
3 &      ABEN\_RD &    39 &  13 & \quart{31}{13}{39}{100} \\
4 &      CART &    37 &  21 & \quart{25}{21}{37}{100} \\
4 &      CART\_RD &    36 &  23 & \quart{25}{23}{36}{100} \\ 
4 &      SVR &    30 &  11 & \quart{25}{11}{30}{100} \\
5 &      ATLM &    -107 &  99 & \ofr \\
    \hline
    
  \nm{miyazaki}\\
  \rowcolor{blue!10} 1 &      CART\_FLASH &    53 &  13 & \quart{46}{13}{53}{100} \\
  \rowcolor{blue!10} 1 &      CART\_DE &    53 &  14 & \quart{46}{14}{53}{100} \\
  \rowcolor{blue!10}1 &      LP4EE &    52 &  11 & \quart{47}{11}{52}{100} \\
  \rowcolor{blue!10} 1 &      ATLM &    50 &  9 & \quart{45}{9}{50}{100} \\
2 &      ABEN\_DE &    46 &  18 & \quart{39}{18}{46}{100} \\
2 &      RF &    46 &  19 & \quart{37}{19}{46}{100} \\
2 &      ABE0 &    45 &  15 & \quart{38}{15}{45}{100} \\  
3 &      CART\_RD &    42 &  20 & \quart{36}{20}{42}{100} \\
3 &      ABEN\_RD &    42 &  24 & \quart{34}{24}{42}{100} \\
3 &      SVR &    41 &  12 & \quart{37}{12}{41}{100} \\    
3 &      CART &    41 &  21 & \quart{31}{21}{41}{100} \\
  \hline

  \end{tabular}
}}

\end{table}


\subsection{Observations}

 Table~\ref{tbl:runtime} shows the runtimes (in minutes) for one of our N*M experiments for each dataset.   From 
 the last column of that table,   we see that the median to maximum
runtimes per dataset range are:
\bi
\item   24 to 54 minutes, for one-way; 
\item   Hence 8 to 18 hours, for the 20 repeats of our N*M experiments.\ei
Performance scores for all datasets are shown in Table~\ref{table_mre}
and Table~\ref{table_sa}. We observe that ATLM and LP4EE performed as expected. Whigham et al.~\cite{Whigham:2015} and Sarro et al.~\cite{SarroTOSEM2018} designed these methods to serve as baselines against which other treatments can be compared. Hence, it might be
expected that  in some cases these   methods will     perform comparatively below other methods.  This was certainly the
case here--
 as seen in  Table~\ref{table_mre}
and Table~\ref{table_sa}, these baseline methods are     top-ranked in 8/18 datasets.

Another thing to observe in Table~\ref{table_mre}
and Table~\ref{table_sa} is that  random search (RD)  also performed as expected; i.e. it was never top-ranked. This is a gratifying result since if random otherwise, then that tend to negate the value of  hyperparameter optimization.

We also see in Table~\ref{table_mre} empirical evidence many of our methods
achieve 
human-competitive results. 
Molokken and Jorgensen~\cite{Jorgensen03}'s
survey of current industry practices reports that human-expert predictions of project effort lie within 30\% and 40\% of the true value; i.e. \mbox{MRE $ \le 40$\%}. Applying that range to Table~\ref{table_mre} we see that
in 6/9 datasets, the best estimator has \mbox{MRE $ \le 35$\%}; i.e. 
they lie comfortably within the stated human-based industrial thresholds.
Also, in a further  2/9 datasets, the best estimator has  \mbox{MRE $ \le 45$\%}; i.e.
they are close to the performance of humans.

The exception to the results in the last paragraph is isbg10 where the best estimator has an 
\mbox{MRE $ = 59$\%}; i.e. our best performance is nowhere close to that
of human estimators. In future work, we recommend researchers use isbg10
as a ``stress test'' on new methods.

\subsection{Answers to Research Questions}

Turning now to the research questions listed in the introduction:

 \noindent{\bf RQ1: Is it best just to use the ``off-the-shelf'' defaults?}
 
 As mentioned in the introduction, 
 Arcuri \& Fraser note that for
 test case generation,   using the default settings
can work just as well as anything else. 
 We can see some evidence of this effect in  Table~\ref{table_mre}
and Table~\ref{table_sa}. Observe, for example, the
  kitchenham results where the untuned ABE0 treatment achieves \colorbox{blue!10}{{\em Rank=1}}.  
  
However,  overall, Table~\ref{table_mre}
and Table~\ref{table_sa} is negative on the use of default settings.
For example,  in  datasets ``albrecht'', ``desharnais'', ``finnish'',
not even one treatments that use the default found in \colorbox{blue!10}{{\em Rank=1}}. 
Overall, if we always used just {\em one} of the methods using defaults (LP4EE, ATLM, ABE0) then that would
achieve best ranks in 8/18 datasets.

 Another aspect to note in the Table~\ref{table_mre}
and Table~\ref{table_sa} results
 are the large differences in performance scores
 between the best and worst treatments (exceptions:   miyazaki's MRE and SA scores do not vary much; and neither does  isbg10's SA scores). That is, there is much to be gained by using the \colorbox{blue!10}{{\em Rank=1}} treatments and deprecating the rest.
 
 In summary,  using the defaults is recommended only in a part of
 datasets. Also, in terms of better test scores,
 there is much to be gained   from tuning. Hence:
 
 \begin{result}{1}
``Off-the-shelf'' defaults
 should be deprecated.
 \end{result}

\noindent{\bf RQ2: Can we replace the old   defaults
 with new defaults?}
 
 If the hyperparameter tunings found by this paper
 were nearly always the same, then this study
 could conclude by recommending better values
 for default settings. This would
 be a most convenient result since, 
 in future when new data arrives, the complexities of this study 
 would not be needed.

\begin{table*}[!b]
\caption{Tunings discovered by hyperparameter selections
(CART+DE). Table rows
sorted by number of rows in data sets
(smallest on top).
Cells in this table show the percent of times a particular choice was made. White text on black denotes choices made in more than 50\% of tunings.
}\label{table:para_dist}
\begin{center}
\footnotesize

\resizebox{1\linewidth}{!}{ 
\begin{tabular}{ r|c|c|c|c|c|c}

~ & \%max\_features & max\_depth & min\_sample\_split & min\_samples\_leaf  \\  
~ & (selected at random; & (of trees) & (continuation & (termination \\ 
~ & 100\% means ``use all'') &   & criteria) & criteria)   \\\cline{2-7} 
~ &
\makecell[l]{
\ 25\%\ 50\%\ 75\%\ 100\%} &
\makecell[l]{
\ $\leq$03 \ $\leq$06 \ $\leq$09 \ $\leq$12} &
\makecell[l]{
\ $\leq$5 \ $\leq$10 \ $\leq$15 \ $\leq$20} &
\makecell[l]{
\ $\leq$03 \ $\leq$06 \ $\leq$09 \ $\leq$12} 
 \\
\hline

kemerer
&\dbox{17}\dbox{34}\dbox{24}\dbox{25}
&\wbox{56}\dbox{36}\dbox{08}\dbox{00}
&\wbox{94}\dbox{03}\dbox{03}\dbox{00}
&\wbox{90}\dbox{03}\dbox{05}\dbox{02}
\\
albrecht
&\dbox{17}\dbox{23}\dbox{19}\dbox{41}
&\wbox{61}\dbox{26}\dbox{11}\dbox{02}
&\wbox{65}\dbox{33}\dbox{02}\dbox{00}
&\wbox{82}\dbox{14}\dbox{04}\dbox{00}
\\
isbsg10
&\dbox{16}\dbox{36}\dbox{24}\dbox{24}
&\wbox{54}\dbox{32}\dbox{10}\dbox{04}
&\dbox{49}\dbox{22}\dbox{15}\dbox{14}
&\wbox{57}\dbox{25}\dbox{11}\dbox{04}
\\
finnish
&\dbox{09}\dbox{04}\dbox{31}\wbox{56}
&\dbox{36}\wbox{52}\dbox{12}\dbox{00}
&\wbox{71}\dbox{16}\dbox{08}\dbox{05}
&\wbox{77}\dbox{18}\dbox{05}\dbox{00}
\\
miyazaki
&\dbox{11}\dbox{23}\dbox{26}\dbox{40}
&\dbox{26}\dbox{49}\dbox{20}\dbox{05}
&\dbox{36}\dbox{27}\dbox{20}\dbox{17}
&\wbox{76}\dbox{15}\dbox{07}\dbox{02}
\\
maxwell
&\dbox{09}\dbox{16}\dbox{37}\dbox{38}
&\dbox{19}\wbox{55}\dbox{21}\dbox{05}
&\dbox{46}\dbox{21}\dbox{18}\dbox{15}
&\wbox{52}\dbox{32}\dbox{14}\dbox{04}
\\
desharnais
&\dbox{21}\dbox{28}\dbox{31}\dbox{20}
&\dbox{47}\dbox{41}\dbox{10}\dbox{02}
&\dbox{33}\dbox{26}\dbox{12}\dbox{29}
&\dbox{35}\dbox{21}\dbox{23}\dbox{21}
\\
kitchenham
&\dbox{08}\dbox{15}\dbox{31}\wbox{46}
&\dbox{07}\dbox{34}\dbox{43}\dbox{16}
&\dbox{41}\dbox{37}\dbox{13}\dbox{09}
&\dbox{43}\dbox{31}\dbox{16}\dbox{10}
\\
china
&\dbox{00}\dbox{06}\dbox{27}\wbox{67}
&\dbox{00}\dbox{02}\dbox{23}\wbox{75}
&\wbox{51}\dbox{36}\dbox{09}\dbox{04}
&\wbox{64}\dbox{30}\dbox{05}\dbox{01}
\\\end{tabular}}

\mbox{KEY: \colorbox{black!10}{\bf 10}\colorbox{black!20}{\bf 20}\colorbox{black!30}{\bf 30}\colorbox{black!40}{\bf 40}\colorbox{black!50}{\bf \textcolor{white}{50}}\colorbox{black!60}{\bf \textcolor{white}{60}}\colorbox{black!70}{\bf \textcolor{white}{70}}\colorbox{black!80}{\bf \textcolor{white}{80}}\colorbox{black!90}{\bf \textcolor{white}{90}}\colorbox{black}{\bf \textcolor{white}{100}}}\%

\end{center}
\end{table*}

Unfortunately, this turns out not to be the case.
Table~\ref{table:para_dist} shows the percent frequencies with which
some tuning decision appears in our {\em M*N-way} cross validations
(this table uses results from DE tuning CART since, as shown below,
this usually leads to best results).
Note that in those results it it not true that across most datasets there is a setting that is usually selected
(thought min\_samples\_leaf less than 3 is often a  popular setting).
Accordingly, we  say that Table~\ref{table:para_dist} shows that there is
much variations of the best tunings. 
Hence, for effort estimation:

 \begin{result}{2}
 Overall, there are no ``best'' default settings.
 \end{result}

Before going on, one curious aspect of the Table~\ref{table:para_dist} results are the 
\%max\_features results; it was rarely most useful to use all features. Except for  finnish and china), best results were often obtained after discarding (at random) a quarter to three-quarters of the features. This is a clear indication that, in future work, it might be advantageous to explore more feature selection for CART models.

\noindent{\bf RQ3: Can we avoid slow hyperparameter optimization?}

\RED
Some methods in our experiments (ABEN\_RD and ABEN\_DE) are slower than others, even with the same number of evaluations, as shown in Table~\ref{tbl:runtime}. Is it possible to avoid such slow runtimes?

Long and slow optimization times are recommended
when their exploration leads to better solutions.
Such better solutions from  slower optimizations
are rarely found in Table~\ref{table_mre}
and Table~\ref{table_sa} (only in 2/18 cases: see the ABEN\_DE results
for kitchenham, and china). Further, the size of the improvements
seen with the slower optimizers   over the best Rank=2 treatments is  small. 
Those improvements come at  runtime cost (in Table~\ref{tbl:runtime}), the  slower optimizers
are one  orders of magnitude slower than other methods). Hence we say that for effort estimation:
\BLACK

\begin{result}{3}
Overall,  our  slowest  optimizers  perform  no
better than  faster ones.
 \end{result}
 
\noindent{\bf RQ4: What  hyperparatmeter optimizers to use for effort estimation?}

 When we discuss this work with our industrial colleagues, they want to know ``the bottom line''; i.e. what they should use or, at the very least, what they should not use. This section offers that advice. We stress that this  section is  based on the above results so, clearly these   recommendations are something that would need to be revised whenever new results come to hand.

\RED

 Based on the above we can assert that
 using  {\em all} the estimators mentioned above is not recommended (to say the least):
 \bi\item
 For one thing, many of them never appear in our top-ranked results.
 \item For another thing, testing all of them on new datasets would be needlessly expensive. Recall our rig: 20 repeats over the data where each of those repeats include slower estimators shown in Table~\ref{tbl:runtime}. 
 As seen in that figure, the median to maximum runtimes for such an analysis for a single dataset
 would take  8 to 18 hours (i.e. hours to days).
\ei

\begin{wraptable}{r}{2.3in}
\vspace{-15pt}
\small
 \caption{Methods ranking of total winning times ({\em Rank=1}), in all 18 experiment cases (9 datasets for both MRE and SA).}\label{tbl:methodsrk}
\centering
{  
\begin{tabular}{clc} \\
  {\textbf{Rank}}& \textbf{Method} & \textbf{Win Times}\\
  \hline
    1 &      CART\_FLASH    &    16/18\\\hline 
    2 &      CART\_DE       &    14/18\\\hline 
    3 &      LP4EE          &    7/18\\\hline 
    4 &      ATLM           &    3/18\\\hline 
    5 &      ABEN\_DE       &    2/18\\
    5 &      ABE0           &    2/18\\\hline 
    6 &      CART           &    0/18\\
    6 &      ABEN\_RD       &    0/18\\
    6 &      RF             &    0/18\\
    6 &      CART\_RD       &    0/18\\
    6 &      SVR            &    0/18\\
    \hline   
 
 \end{tabular}}
\vspace{-10pt}
\end{wraptable}

Table~\ref{tbl:methodsrk} lists the best that can be expected if an engineer  chooses one of the
estimators in our experiment, and applied it to all our datasets.
The fractions shown at right come from counting optimizer frequencies in the top-ranks of  Table~\ref{table_mre}
and Table~\ref{table_sa}. Note that
the champion in our experiment is ``CART\_FLASH'', which ranked as `1' in 16 out of all 18 cases. One close runner-up is ``CART\_DE'', which has 2 cases less in number of winning times. Those two estimators usually have good performance among most cases in the experiment.

Beside the two top methods, none of the rest estimators could reach even half of all cases. Including those untuned baseline methods  ($\mathit{ATLM}$,  $\mathit{LP4EE}$).
Hence, we cannot endorse their use for generating estimates to be shown to business managers.
That said, we do still endorse their use as a  baseline methods, for methodological reasons in effort estimation research (they are useful for generating a quick result against which we can compare other, better, methods).

Hence, based on the results of  Table~\ref{tbl:methodsrk}, 
for similar effort estimation tasks, we recommend:




\BLACK  
\begin{result}{4}
For new datasets, try a combination of {\em CART} with the optimizers {\em differential evolution} and {\em FLASH}.
 \end{result}

\section{Threats to Validity}\label{sect:threats}
 \textbf{Internal Bias:} Many of  our methods contain stochastic random operators. To reduce the bias from random operators, we 
repeated our experiment in 20 times and applied statistical tests to remove spurious distinctions.

 \textbf{Parameter Bias:} For other studies, this is a significant question
 since (as shown above) the settings to the control parameters of the learners
 can have a positive effect on the efficacy of the estimation.
 That said, 
 recall that much of the technology of this paper concerned methods to explore
 the space of possible parameters.
 Hence we assert that 
 this study suffers much less parameter bias than other studies.
 

\textbf{Sampling Bias:} While we tested OIL on the nine datasets, it would be inappropriate to conclude that OIL tuning  always perform better than
others methods for all datasets.
As researchers, what we can do to mitigate this problem is to carefully document out method, release out code,
and encourage the community to try this method on more datasets, as the occasion arises.

\section{Related Work}

In software engineering, hyperparameter optimization techniques have been applied to some sub-domains, but yet to be adopted in many others. One way to characterize this paper is an attempt to adapt recent work in hyperparameter optimization in
software defect prediction to effort estimation.  Note that, like in defect prediction, this article has also concluded that Differential Evolution is an useful method.

\begin{figure}[ht]
\caption{Literature review of hyperparameters tuning on 52 top defect prediction papers \cite{fu2016differential}}\label{fig:litfigure}
\vspace{10pt}
\centering
\includegraphics[width=0.75\textwidth]{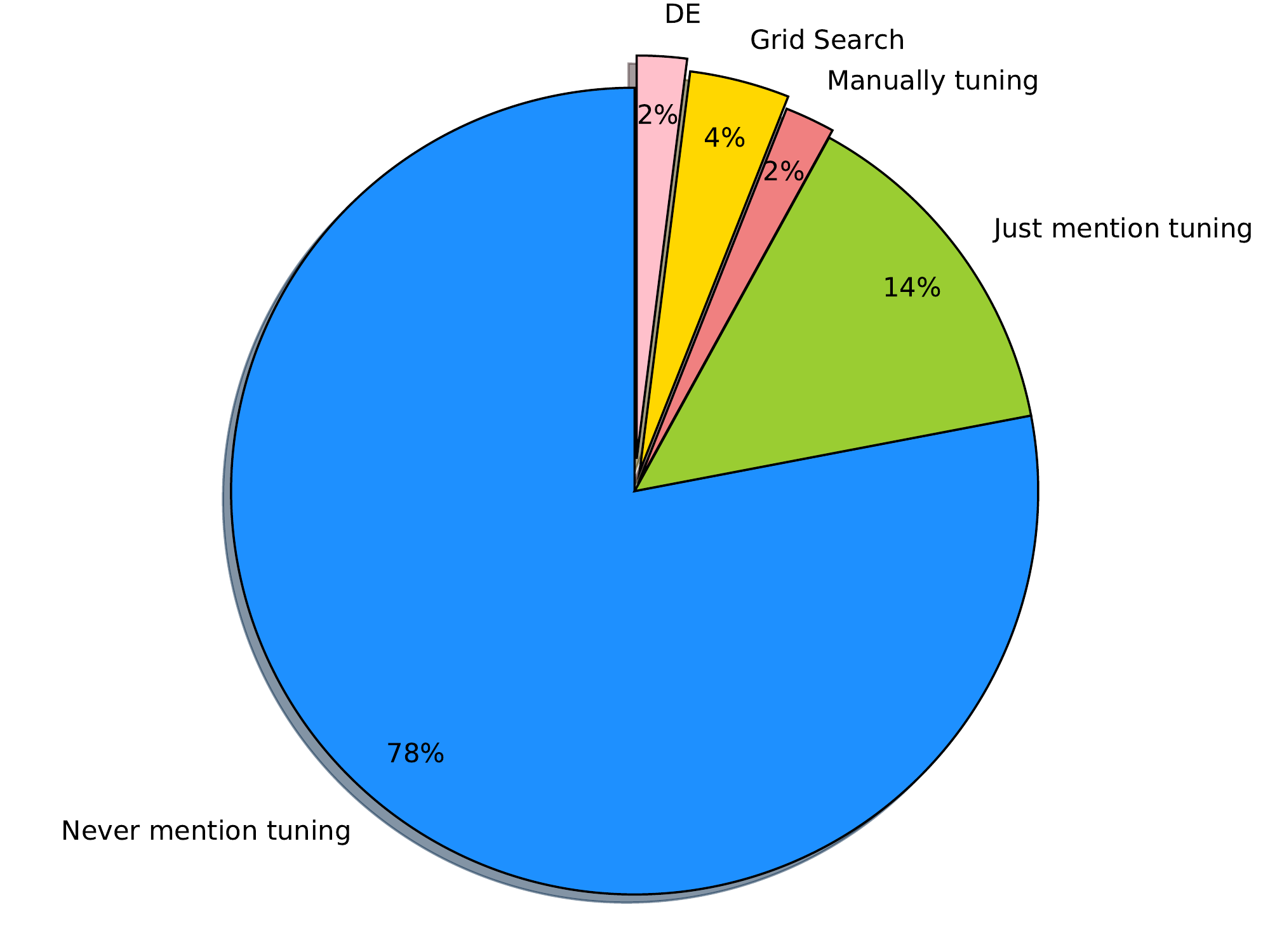}
\vspace{-0.65cm}
\end{figure}

Several SE defect prediction techniques rely on static code attributes~\cite{krishna16bellwether, nam15hetero, tan15online}. 
Much of that work has focused of finding and employing complex and ``off-the-shelf'' machine learning models~\cite{menzies07defect, moser08defect, elish08defect}, without any hyperparameter optimization. According to a literature review  done by Fu et al.~\cite{fu2016differential}, as shown in Figure \ref{fig:litfigure}, nearly 80\% of highly cited papers in defect prediction
do not mention   parameters tuning (so they  rely on  the default parameters setting of the predicting models).

Gao et al.~\cite{gao2011choosing} acknowledged the impacts of the parameter tuning for software quality prediction. For example, in their study, ``{\em distanceWeighting}'' parameter was set to ``{\em Weight by 1/distance}'', the {\em KNN} parameter ``{\em k}'' was set to ``30'', and the ``{\em crossValidate}'' parameter was set to ``{\em true}''. However, they did not provide any further explanation about their tuning strategies.

As to methods of tuning,
Bergstra and Bengio~\cite{Bergstra:2012} comment that
{\em grid search}\footnote{For $N$ tunable  option, run $N$ nested for-loops to explore their ranges.}  
is very popular since
 (a) such a simple search to gives researchers some degree
of insight; (b) grid search has very  little technical overhead for its implementation; (c)  it is simple to automate and parallelize; (d) on a computing
cluster, it can find better tunings than sequential optimization (in
the same amount of time). 
That said, Bergstra and Bengio deprecate grid search since that style of search is not more effective than more randomized searchers if the underlying search space is inherently low dimensional. \RED This remark is particularly relevant
to effort estimation since datasets in this domain are
often low dimension~\cite{kocaguneli2013active}. \BLACK

Lessmann et al.~\cite{Lessmann08} used grid search to tune parameters as part of
their extensive analysis of different algorithms for defect prediction. However, they only tuned a small set of their learners while they used the default settings for the rest. Our conjecture is that the overall cost of their tuning was too expensive so they chose only to tune the most critical part.

Two recent studies about investigating the effects of parameter tuning on defect prediction were conducted by Tantithamthavorn et al.~\cite{tanti16defect, tanti18} and Fu et al.~\cite{Fu2016TuningFS}. Tantithamthavorn et al. also used grid search while Fu et al. used differential evolution. Both of the papers concluded that tuning rarely makes performance worse across a range of performance measures (precision, recall, etc.). Fu et al.~\cite{Fu2016TuningFS} also  report that different datasets require different hyperparameters to maximize performance. 

One major difference between the studies of Fu et al.~\cite{Fu2016TuningFS} and Tantithamthavorn et al.~\cite{tanti16defect} was the computational costs of their experiments. Since Fu et al.'s differential evolution based method had a strict stopping criterion, it was significantly faster.

Note that there are several other methods for hyperparameter optimization and we aim to explore several other method as a part of future work. But as shown here, it requires much work to create and extract conclusions from a hyperparameter optimizer. One goal of this work, which we think we have achieved, to identify a simple baseline method against which subsequent work can be benchmarked.

\section{Conclusions and Future Work} \label{sect:conclusion}

Hyperparameter optimization is known to  improve the performance of many software analytics tasks such as software defect prediction or 
 text classification~\cite{agrawal2017better, AGRAWAL2018, Fu2016TuningFS, tanti18}.
Most prior work in this effort estimation optimization only explored very small datasets~\cite{li2009study} or used
estimators that are not representative of  the state-of-the-art~\cite{Whigham:2015,SarroTOSEM2018}.
Other researchers assume that the effort model is a specific parametric form (e.g. the COCOMO equation), which greatly limits the amount of data that can be studied.
Further, all that prior work needs to be revisited given the existence of recent and very prominent
methods; i.e. ATLM from TOSEM'15~\cite{Whigham:2015} and LP4EE from TOSEM'18~\cite{SarroTOSEM2018}.

Accordingly, this paper conducts a thorough investigation of     hyperparameter   optimization for effort estimation
using methods (a)~with no   data
feature assumptions (i.e. no COCOMO data);
(b)~that
 vary
many  parameters (6,000+ combinations);
that    tests  its results   on 9 different sources with data on 945 software projects; 
(c)~which  uses optimizers   representative of the  state-of-the-art 
(DE~\cite{storn1997differential}, FLASH~\cite{nair2017flash});
and which 
(d)~benchmark results 
against  prominent methods such as 
  ATLM and LP4EE.

These results were assessed with respect to the  Arcuri and  Fraser's concerns mentioned in the introduction; i.e. sometimes
hyperparamter optimization can be both too slow and not effective.
Such pessimism    may indeed apply  to  the  test  data  generation  domain.
However, 
the results of this paper show that 
 there  exists
other domains like effort estimation where hyperparameter
optimization  is  both
useful
and
fast. After  applying  hyperparameter optimization,  large  improvements  in  effort
estimation  accuracy  were  observed  (measured  in  terms  of  the
 standardized  accuracy).
From those results, we can recommend using a combination of regression trees
(CART)  tuned  by  different  evolution and FLASH.  This
particular combination of learner and optimizers can achieve
in  a few minutes  what  other  optimizers  need  longer runtime
of  CPU  to  accomplish.


\RED This study is a very extensive explorations of hyperparameter optimization and effort estimation yet undertaken. There are still very many
options not explored here. Our current
plans for future work include the following:

\BLACK
\bi
\item Try other learners:
e.g. neural nets, bayesian learners or AdaBoost;
\item Try other data pre-processors.
We mentioned above how it was curious that
max features was often less than 100\%.
This is a clear indication that, we
might be able to further improve our estimations results by adding
more intelligent feature selection to, say, CART.
\item Other optimizers. For example,
combining DE and FLASH might be a fruitful way
to proceed.
\item Yet another possible future direction
could be hyper-hyperparamter optimization. In
the above, we used optimizers like differential
evolution to tune learners. But these optimizers
have their own control parameters. Perhaps there 
are better settings for the optimizers? Which could be found via hyper-hyperparameter optimization?
\ei
Hyper-hyperparameter optimization could be a very slow
process. Hence, results like this paper could be
most useful since here we have identified
optimizers that are
very fast and very slow (and the latter would
not be suitable for hyper-hyperparamter optimization).

In any case,
we  hope  that
OIL  and the results of this paper will  prompt  and  enable
more research on better methods to tune software effort
estimators.
To that end, we have placed our scripts and data online at https://github.com/arennax/effort\_oil\_2019

\section*{Acknowledgement}
This work was partially funded by a National Science Foundation
Award 1703487.

\bibliographystyle{spbasic}


\end{document}